\documentclass{iaus}
\usepackage{graphicx}

\newcommand\etal{\mbox{\textit{et al.}}}

\title[Bulges of lenticulars] %% give here short title %%
{Age and metallicity of the bulges in lenticular galaxies}

\author[Silchenko]   %% give here short author list %%
{Olga K. Sil'chenko$^1$}%
\affiliation{$^1$Sternberg Astronomical Institute, Moscow 119991, Russia
\break email: olga@sai.msu.su \\[\affilskip]
}

\pubyear{2007}
\volume{245}  %% insert here IAU Symposium No.
\pagerange{xxx--yyy}
\date{?? and in revised form ??}
\jname{Formation and Evolution of Galaxy Bulges}
\editors{M. Bureau et al., eds.}
\begin{document}

\maketitle

\begin{abstract}
Panoramic spectroscopy of the sample of 80 nearby lenticular galaxies
is presented. The SSP-equivalent ages, [Z/H], and [Mg/Fe] are determined
through the Lick indices H$\beta$, Mgb, and $\langle \mbox{Fe} \rangle$
separately for the nuclei and for the bulges. About a half of the sample
contain chemically distinct nuclei, more metal-rich and
younger than the bulges. The statistics of stellar population properties
for the nearby S0s is discussed.
\end{abstract}

\firstsection % if your document starts with a section,
\section{Introduction}

Lenticular galaxies are perhaps the only class of nearby galaxies
which is certainly formed rather recently. Direct observations of galaxy
clusters at various redshifts have shown that the fraction of S0s in
clusters has risen from 0--10\%\ to 50\%--60\%\ between the lookback times
of 6 and 2 Gyr (\cite[Fasano et al. 2000]{fasano},
\cite[Desai et al. 2007]{desai_s0}).  The common point
of view is that S0s galaxies are
transformed from spirals due to some external action related to
dense environment and/or hot intergalactic medium. Theoretical
considerations propose a lot of mechanisms for this transformation: ram
pressure by hot intracluster medium (\cite[Quilis et al. 2000]{quilis}),
tidal stripping and heating (\cite[Larson et al. 1980]{ltcs0},
\cite[Byrd \& Valtonen 1990]{byrdval90}),
harassment (\cite[Moore et al. 1996]{moore96}),
minor merger... Probably, different mechanisms may play role in different
environment types, because in fact you can meet lenticular galaxies
in any environments, from field to clusters. An interesting thing is that
many of these mechanisms lead to gas concentration in the very centers
of the galaxies at the moment of their transformation. It means that
we can expect  secondary nuclear star formation bursts and intermediate-age
stellar populations in the nuclei of nearby S0s. Also, the statistics on
bulge-to-disk luminosity ratios show that nearby S0s have on average more 
massive bulges than nearby spirals (e.g.
\cite[Trujillo et al. 2002]{trujill02}), so their
transformation must include some process of bulge growth. It means that 
the bulges must have suffered some rejuvenation during the last 5 Gyr, 
and their stellar populations may be on average younger than those in
nearby ellipticals. So there are good reasons to extract S0 galaxies from
the wider samples of `early-type' and `red-sequence' galaxies and
to search for particular properties of their stellar populations.

\section{Sample}

We consider a sample of 80 nearby S0s in the wide range of luminosity.
Our sample is not complete but rather representative. We have undertaken a
retrieval over the HYPERLEDA with the following restrictions:

\begin{itemize}

\item{$v_r < 3000$ km/s;}
\item{$-3 \le t \le 0$, where $t$ is the numerical morphological-type indicator;}
\item{$\delta (2000.0) >0$;}
\item{$B_T^0 \le 13.0$.}

\end{itemize}

Strong Seyfert nuclei or nuclear star bursts were
excluded. The search gave us a list of 148 S0s, with 50 Virgo
members among them. We have observed 66 targets from
this list, including 10 Virgo members.
Also a few more distant luminous galaxies as well as nearby
fainter ones were added to the sample to expand the range of luminosities.

By following NOG survey of groups (\cite[Giuricin et al. 2000]{nog}),
we put our galaxies into four types of environments: those
belonging to clusters (Virgo and Ursa Major), group central galaxies 
(the brightest galaxies in groups), group second-rank members, and field
galaxies. Finally, we have 13, 22, 33, and 12 galaxies in every class.

The novelty of our approach to bulges is that we do not use {\it
aperture} spectral data focused onto the centers of galaxies.
Inspired by the panoramic spectroscopy benefits, we consider separately 
unresolved stellar nuclei and bulges taken as rings between 
$R=4^{\prime \prime}$ and $R=7^{\prime \prime}$ ; 
the latter $R$ corresponds to 1.3 kpc for the most distant
galaxies.

\section{Observations and data reduction}

All the observations have been made with the integral-field spectrograph
of the Russian 6-m telescope, Multi-Pupil Field/Fiber
Spectrograph (MPFS). The spectrograph had several modifications.
In 1994--1998 it was a TIGER-mode spectrograph, with the fields
of view of about $10 \times 16$ square elements, of
$1.3^{\prime \prime}$ each. The spectral range was tight, and the
spectral resolution, on average about of 5~\AA, varied strongly over
the field of view. In 1998 this spectrograph was replaced by another MPFS,
which contained fibers transmitting light from the microlenses
to the slit (\cite[Afanasiev et al. 2001]{mpfsref}).
From 2002 upto now we observe with the field of view
of $16 \times 16$ elements, of $1^{\prime \prime}$ each, the spectral
range of 1500~\AA, and the working
spectral resolution of about 3~\AA\ that permits to obtain reliable
kinematical data in addition to the absorption-line indices.

Two spectral ranges are used: the green one, 4200-5700~\AA, which
provides Lick indices and stellar kinematics, and the
red one, 5800-7200~\AA, where strong emission lines
can be found to study ionized-gas kinematics.
We co-added the bulge elementary spectra over the rings
to get index accuracy for the bulges comparable to that for the nuclei.
Stellar velocities and velocity dispersions have been calculated 
by cross-correlation with template star spectra, gas velocities
-- by measuring baricentres of emission lines. The Lick indices
H$\beta$, Mgb, Fe5270, and Fe5335 are measured for every nucleus and
bulge. The Lick index system is calibrated by observing a sample of
standard Lick stars (\cite[Worthey et al. 1994]{woretal}).
Having observed a dozen galaxies
more than once, we have assured that internal precision of the indices
is 0.15~\AA\ for H$\beta$ and iron indices, and 0.1~\AA\ for Mgb.

\section{Results}

Three characteristics of stellar populations, age $T$, global metallicity
[Z/H], and magnesium-to-iron ratio [Mg/Fe], are determined separately
for the nuclei and for the bulges by using the models of old stellar
populations by \cite[Thomas et al. (2003)]{thomod}.
SSP approach (single-burst one-metallicity population) is applied, 
so the parameters determined are close to luminosity-averaged ones.
The age $T$ and metallicity [Z/H] are determined firstly, by confronting 
H$\beta$ to [MgFe]$=(\mbox{Mgb}\langle \mbox{Fe} \rangle)^{1/2}$ 
(insensitive to the Mg/Fe ratio), and then the abundance ratio [Mg/Fe]
is determined by confronting $\langle \mbox{Fe} \rangle$
to Mgb for a given $T$.
The detailed tables can be partly found in my
recent paper (\cite[Sil'chenko 2006]{lenssum}).

\begin{figure*}
\includegraphics[height=4.3cm]{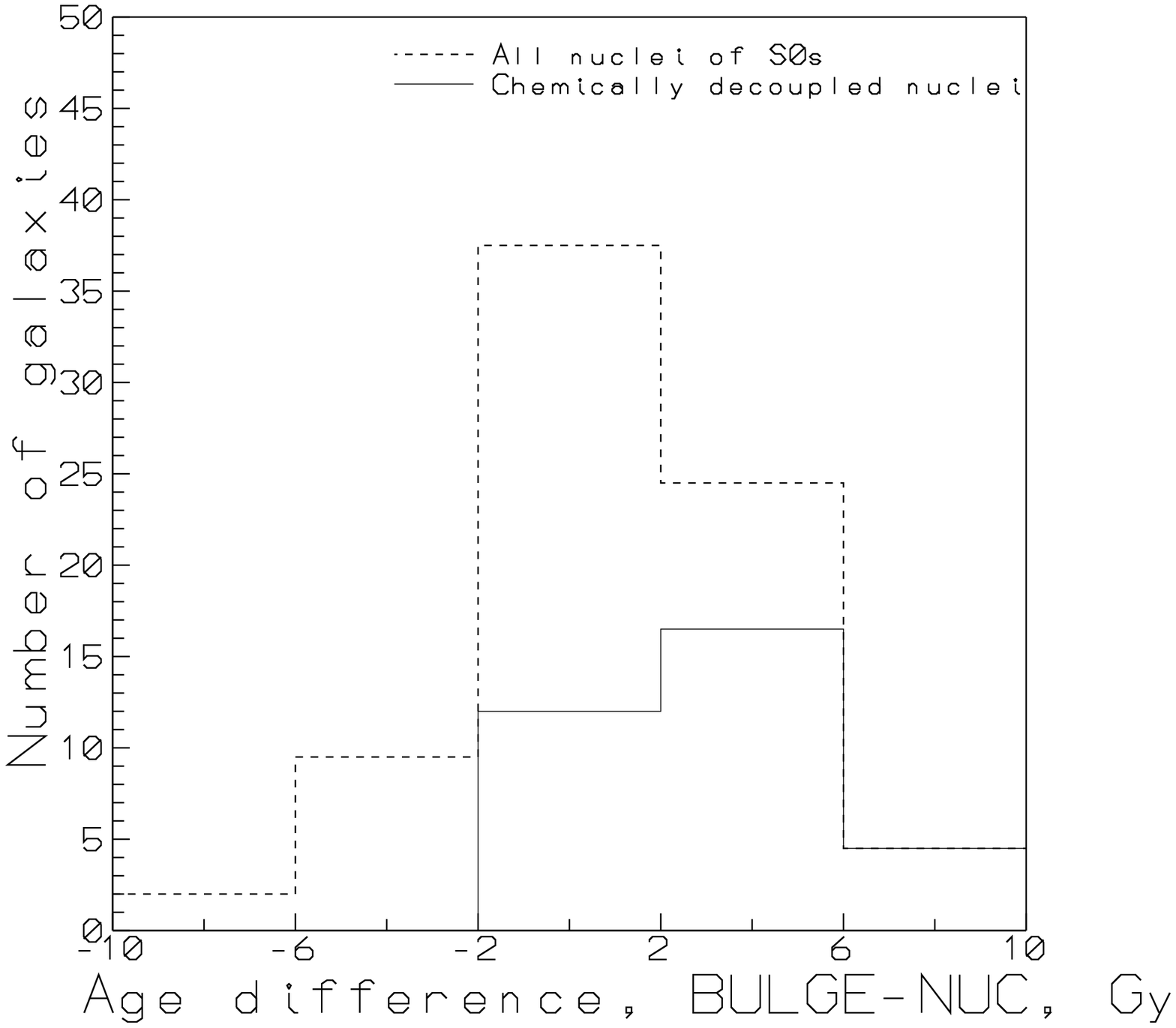}
\includegraphics[height=4.3cm]{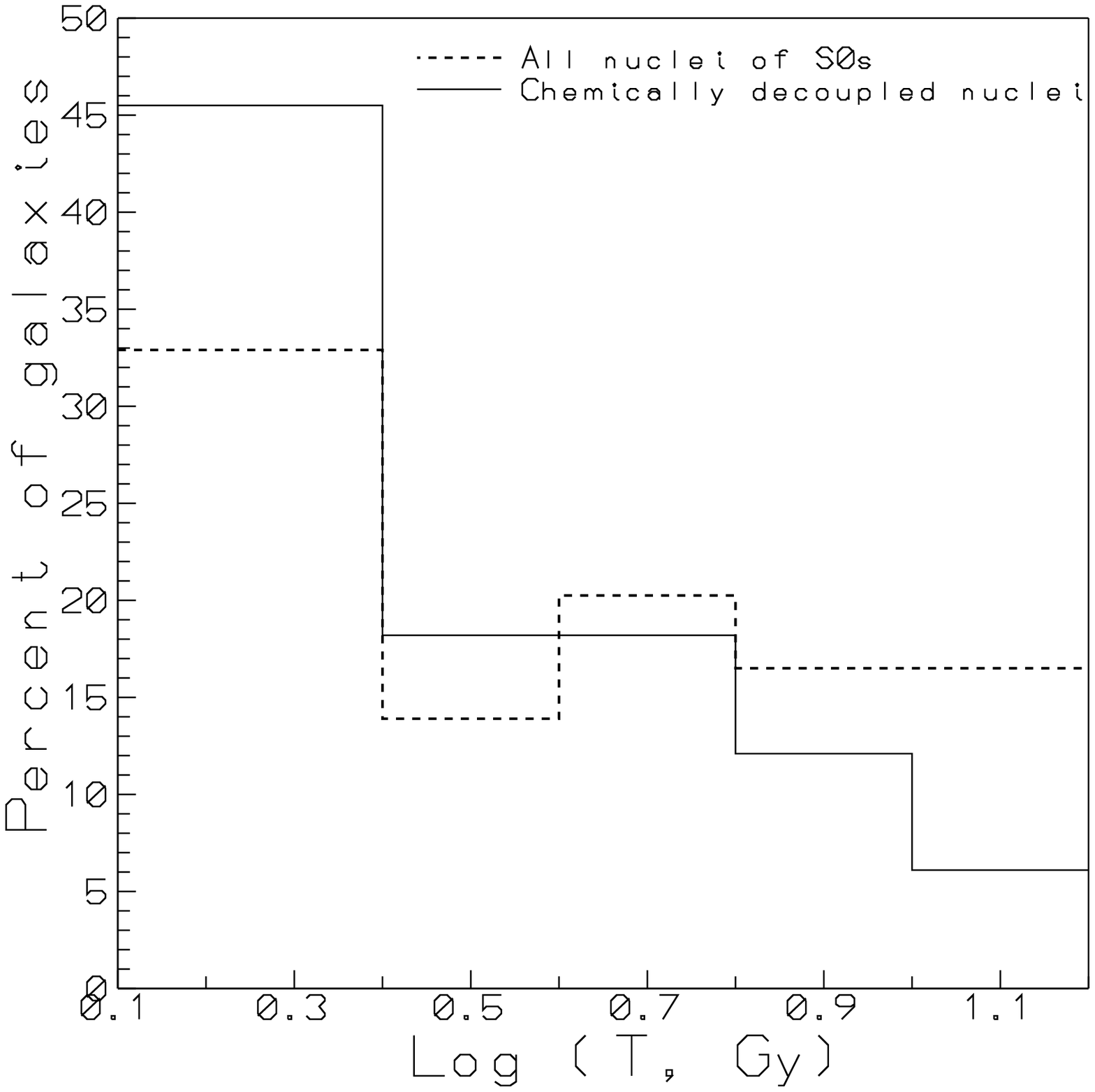}
  \caption{The results on mean stellar ages for the
  chemically distinct nuclei in comparison to the total sample
  of the nuclei in lenticular galaxies.
    (\textit{a}) the histogram of the stellar age difference
    between the nuclei and the bulges; the chemically distinct
    nuclei are {\bf always} younger than the surrounding bulges;
    (\textit{b}) the distributions of the absolute values of the
    SSP-equivalent ages for the nuclei; the chemically distinct
    nuclei are separately shown by a solid line.}
\label{decnucage}
\end{figure*}

By starting our study of nuclei and bulges of early-type galaxies with
the MPFS, we discovered immediately that the (unresolved) nuclei differ
strongly from the surrounding bulges as concerning the absorption-line
indices, namely, the nuclei demonstrated stronger metal absorption lines
(\cite[Sil'chenko et al. 1992]{decnuc}). Now, with our sample of
80 lenticular galaxies observed with the MPFS, we can state that about
half of them have chemically distinct nuclei. If we select the galaxies
where the nuclei are more metal rich than the bulges by $+0.3$ dex
and more (by a factor of 2 and more), we come to the estimate of 42\%.
The limit put by us onto the metallicity difference to
select chemically distinct nuclei is rather conservative and is defined
by the fact that current estimates of the metallicity gradients in
spheroidal galaxies are about 0.2--0.3 dex per radius dex. The chemically
distinct nuclei selected by us have the {\it same} metallicity drop inside
the radius equal to {\it one or two} resolved spatial elements. By comparing
the subsample of the S0s with the chemically distinct nuclei to the
total sample of nearby S0s (Fig.~\ref{decnucage}), we have assured that
the chemically distinct nuclei are {\bf always} younger than the surrounding
bulges: the mean age difference between the chemically
distinct nuclei and their surrounding bulges is 2.8 Gyr 
while the age difference between the nuclei and the bulges
for the whole sample is only 1 Gyr. The age distribution of
the chemically distinct nuclei peaks strongly at $T=2$ Gyr
whereas the total nuclear age distribution is rather
flat between $T=1$ and $T=12$ Gyr.

So it is necessary to separate the nuclei and the bulges before
making any conclusions about the properties of any of them. Nuclei and
bulges have quite different evolution and differ as concerning
their present stellar populations. Further we consider
the bulges as taken in the rings between the radii of $4^{\prime \prime}$
and $7^{\prime \prime}$ -- within the areas of bulge domination
but beyond those affecting by the nuclei influence due to seeing effects.

By observing a representative sample of nearby S0s, we have tried
to cover homogeneously all types of environments. We have more than
ten galaxies in every environment class, so we are able
to reveal any difference of stellar population parameters due to
the environment type. However, at all diagnostic diagrams `index vs
index' the cluster S0s occupy just the same area as the group
center galaxies, and group second-rank members are indistinguishable from
the field galaxies. So we unite all the galaxies in two big groups:
S0s in dense environments, namely, in clusters and group centers, and
S0s in sparse environments, namely, in field and off-centered in groups.
We see some difference between these two groups. In particular,
at the diagram `H$\beta$ vs [MgFe]' which serves for age and metallicity
determination, the bulges of S0s in dense environments are concentrated as
compact point cloud, elongated from the parameter combination
($T=12$ Gyr, [Z/H]$=0.0$ dex) toward the combination of
($T=5$ Gyr, [Z/H]$=+0.3$ dex),
while the bulges of the galaxies in sparse environments are
spread over all the ages, from 1 to 15 Gyr, in the metallicity
band of $0.0- +0.3$ dex. The median stellar ages of the bulges
are 4.0 Gyr in sparse environments and 6.7 Gyr in dense environments;
the corresponding estimates for the nuclei are 2.4 and 4.1 Gyr, so
the nuclei are on average younger than the bulges in any types of
environments.
To test if the age differences due to environment density is real,
we must analyze age correlation with the stellar velocity dispersion.

Such relations are well-studied for elliptical galaxies in numerous
works. We can compare our results obtained for the bulges of S0s,
first of all, with these data. But as
we separate the nuclei and the bulges when studying the properties
of stellar populations, similarly, we wish to measure stellar velocity
dispersion separately in the very centers (`inside the nuclei') and
in the bulges, at $R \sim 5^{\prime \prime}$. Indeed,
the results reveal again distinction between
the nuclei and the bulges: the stellar velocity dispersion in the
nuclei is typically larger than that in the bulges, and sometimes
the difference reaches 70--80 km/s, with the mean of 27 km/s.
To make a fair comparison, we confront the stellar population parameters
to the {\it bulge} stellar velocity dispersion. This reduces our
sample to 52 galaxies, because accurate stellar velocity dispersion
mapping in the full mass range over the full field of view
becomes possible only with the spectral resolution of 3~\AA, after 2002.
Below we analyze the correlations over the range of $\sigma _* =
50 - 200$ km/s.

\begin{figure*}
\includegraphics[width=4.3cm]{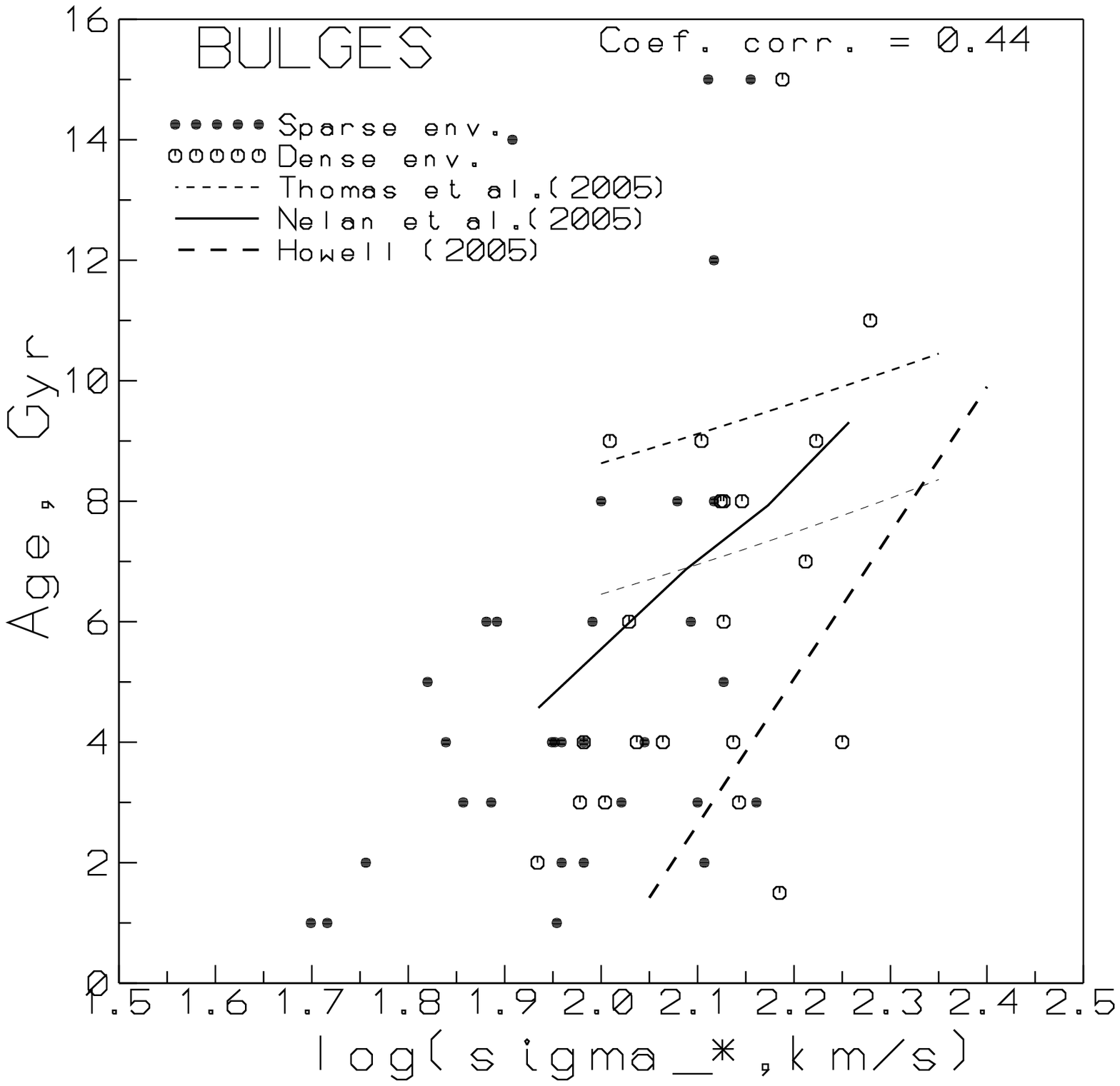}
\includegraphics[width=4.3cm]{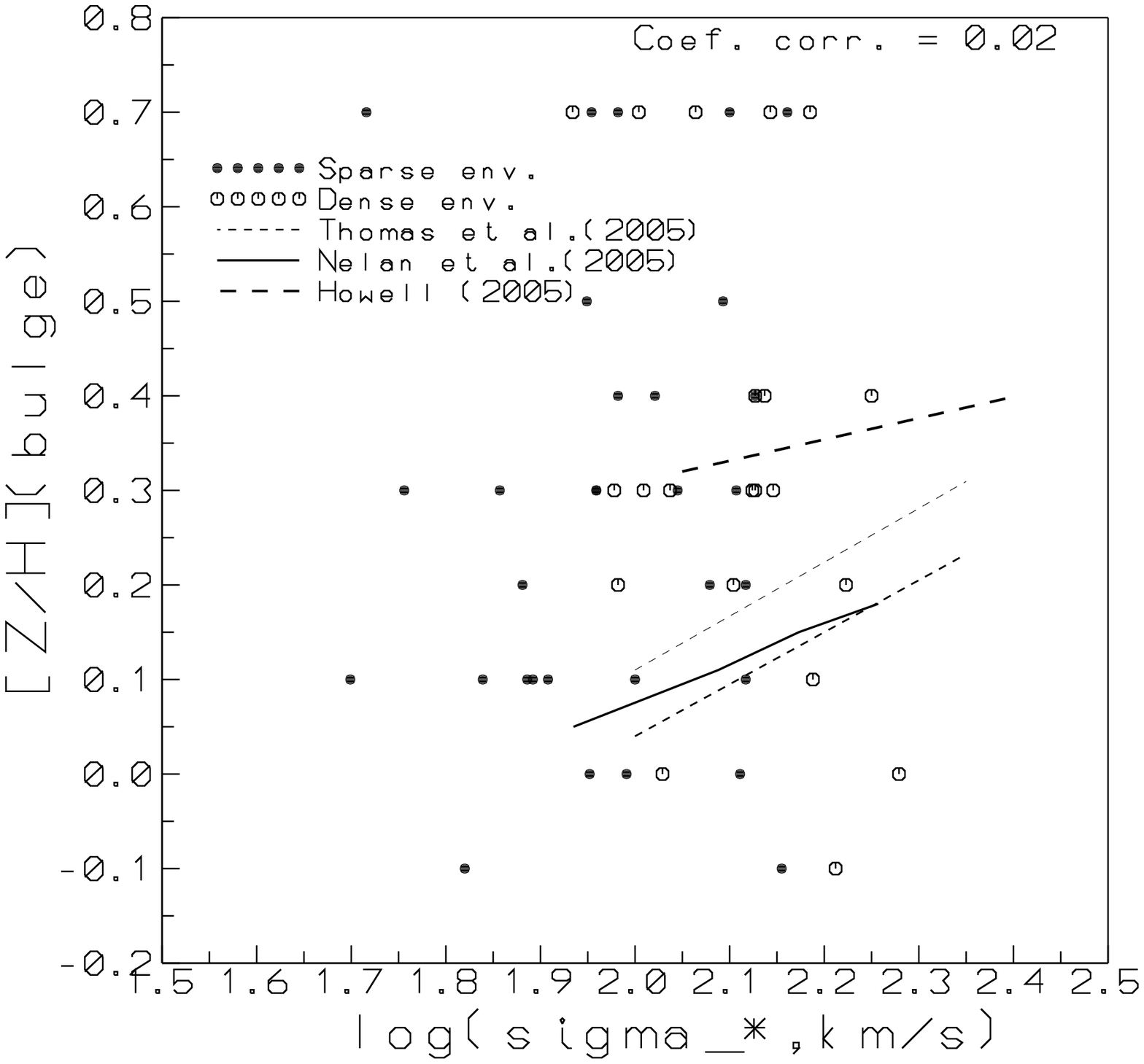}
\includegraphics[width=4.3cm]{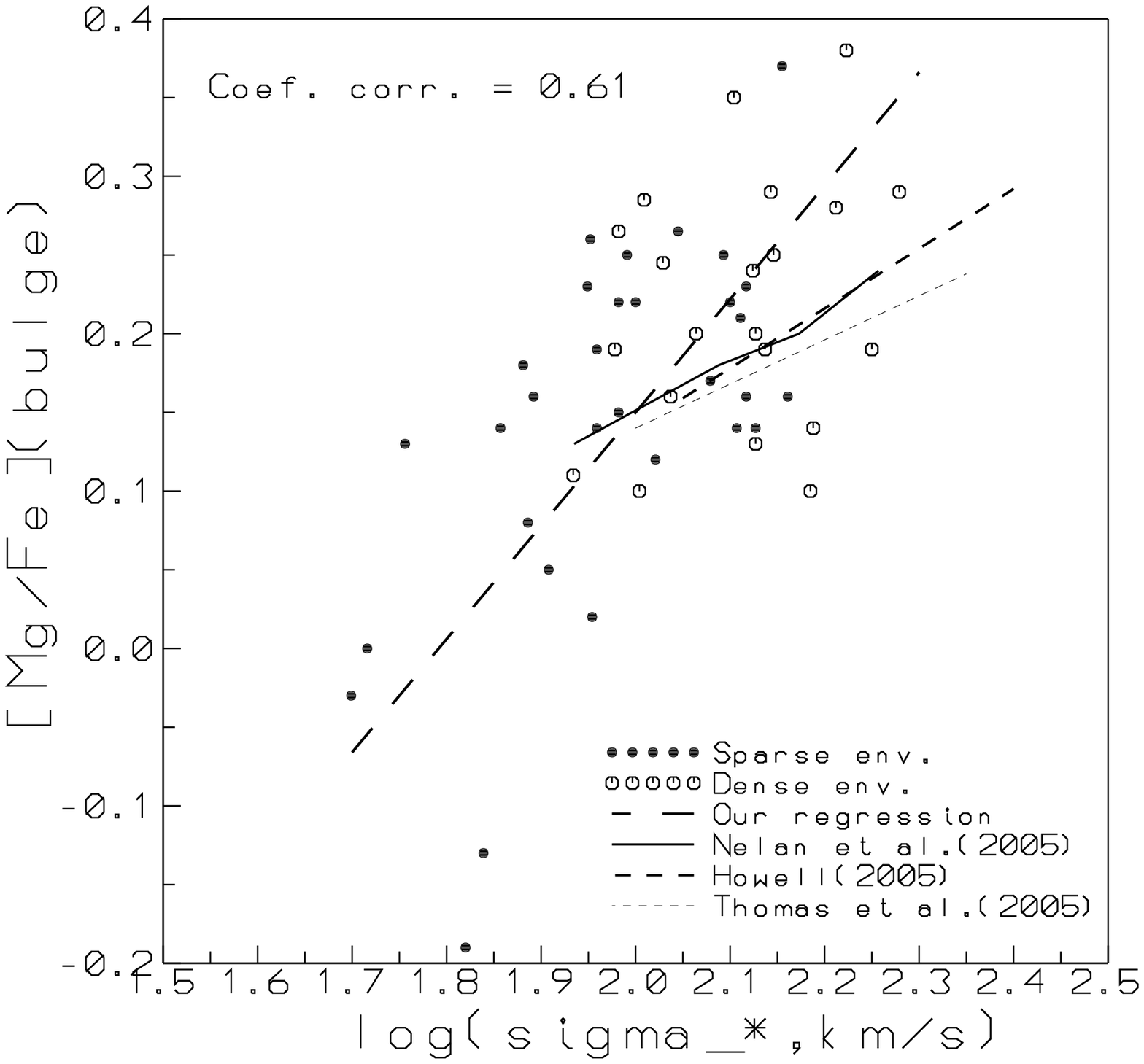}
  \caption{Correlations between the stellar population parameters
and the logarithm of the local stellar velocity dispersion for the
bulges of nearby lenticular galaxies. Local relations for various
samples of early-type galaxies from the literature are also shown.
    (\textit{a}) the SSP-equivalent age vs $\log \sigma _*$;
    (\textit{b}) the total metallicity vs $\log \sigma _*$;
     (\textit{c}) the magnesium-to-iron ratio vs $\log \sigma _*$.}
\label{s0corrs}
\end{figure*}

In Fig.~\ref{s0corrs} we show dependencies of $T$, [Z/H], and [Mg/Fe]
on $\log \sigma _*$, in comparison with the recent results for
early-type galaxies in various types of environment from
\cite[Thomas et al. (2005)]{thogal}, \cite[Nelan et al. (2005)]{nelan},
and \cite[Howell (2005)]{howell1} which have been obtained
for more massive galaxies, with $\sigma _* >100$ km/s typically.
One can see that the age and metallicity dependencies on
$\log \sigma _*$ are consistent with those for elliptical galaxies.
The age correlates with $\log \sigma _*$, and the metallicity
does not (\cite[Howell 2005]{howell1}). However, the correlation of the 
abundance ratio [Mg/Fe] with $\log \sigma _*$
is the strongest, and the slope of the regression is much steeper than
the slope which is consistently found by several groups of investigators
for elliptical galaxies. What does it mean? If we treat the correlation
of [Mg/Fe] with the mass of spheroid as an evidence for more effective
(and brief) star formation in deeper potential well, we may suggest that
the relation found for the bulges of S0s is fundamental for some
early formation process. Consequently, the flatter relation for ellipticals
may be a result of later `dry mergers' which increase the mass of spheroids
leaving the properties of stellar population corresponding to the masses
of smaller progenitors.

\section{Conclusions}\label{sec:concl}

Unresolved stellar nuclei in early-type disk galaxies
have their own evolution: 42\%\ of nearby S0s have
$\Delta$[Z/H](nuc-bul)$=+0.3$ dex and more and
the mean $\Delta T$(nuc-bul)$ = -2.8$ Gyr. The difference of
the stellar velocity dispersion $\Delta \sigma _*$(nuc-bul)
may reach 70 km/s, with the mean of 27 km/s. So to study bulges,
one needs to measure off-nuclear zones.

In sparse environments there is a much larger spread of ages than
in clusters and group centers. However when correlating the ages,
metallicities, and Mg/Fe ratio with the stellar velocity dispersion,
there is no separation due to the environment type.
The SSP-equivalent ages correlate with the local stellar velocity dispersions,
while the SSP-equivalent metallicities do not. But the age spread
rises to larger $\sigma_*$  - unlike ellipticals
(\cite[Caldwell et al. 2003]{caldwell}).
The abundance ratio [Mg/Fe] correlates strongly with the stellar velocity 
dispersion, and the relation slope is much steeper than for ellipticals.

\begin{acknowledgments}
This study has been supported by the grants of the Russian
Foundation for Basic Researches (RFBR), nos. 01-02-16767, 04-02-16987,
and 05-02-19805-MFa. My attendance at the Symposium is due to the
IAU grant and to the RFBR travel grant no. 07-02-08314.
\end{acknowledgments}

\begin{discussion}

\discuss{Pipino}{Why do you think that the bulges of S0s have not
experienced mergers?}

\discuss{Silchenko}{S0 galaxies have large-scale stellar disks,
and these disks are old. It means that recent major merger can
be excluded because they have to destroy stellar disks.}

\end{discussion}

\end{document}